\documentstyle[preprint,aps]{revtex}
\begin{document}
\baselineskip=20pt

\section{Introduction}

{\indent

   It is widely believed that the Universe started by big bang.
 When the temperature dropped to a few hundred MeV, the Universe 
 experienced a phase transition from the quark-gluon plasma (QGP) phase 
 to the hadron 
 phase. If the phase transition was of the first order, the system went 
 through the mixed phase of the quark-gluon and the hadron phases. In 
 this case, the Universe would have become inhomogeneous. Such a 
 non-uniform scenario of the early  Universe was studied  by several 
 authors in terms of the primordial 
 nucleosynthesis and the existence of the strange quark 
 matter,
 which may be a candidate of the dark 
 matter \cite{Witt}
-\cite{Sumi0}.  The condition of forming non-uniform Universe 
 depends largely on the property of the QCD  phase 
 transition as the critical temperature $T_c$ and the surface 
 tension $\alpha$. Although QCD is the theory of quarks and gluons, it is 
 difficult to solve its dynamics to get  such information.

  Hence, we resort to large numerical simulations in the framework 
 of the lattice QCD  theory. The critical temperature 
  and the behavior of the phase transition are investigated by the 
  lattice simulations. 
   In the pure gauge case, the phase transition is of the first order and 
 the critical temperature 
  is $ \,T_c =0.26\, \sim \, 0.28$ GeV \cite{Karsch0,Karsch}. On the other hand, the full QCD 
  simulation provides$\, T_c \sim 0.15$ GeV  and the order of the phase 
  transition is under debate \cite{Kanaya}. 
The surface tension between the confining and deconfining phases
   has been also investigated with the Monte Carlo 
simulation \cite{Kajantie,Huang},
   in particular, on a  large lattice as $36^{2}\times 48\times 
   6$ \cite{iwasaki}. The recent calculations of the surface tension  $\alpha$ 
   provide the surface tension at $\alpha/T_{c}^{3}=0.0292$  for 
   $N_{t}=4$ and $\alpha/T_{c}^{3}=0.0218$ for   $N_{t}=6$, 
   where $N_{t}$ is the value of temporal extents. 
   Taking $T_{c}=0.26$ GeV, we find $\alpha^{1/3}=80$ MeV for 
   $N_{t}=4$ and  $\alpha^{1/3}=73$ MeV  for  $N_{t}=6$.  
   Here, the surface tension obtained in the lattice QCD is the 
value at the thermodynamical critical temperature.
In general, it is hard to get the information on the metastable state
from the lattice simulation, and therefore the upper and the lower 
critical temperatures \cite{Ichie}, 
for instance, cannot be obtained by the lattice QCD. 

On the other hand, the dual Ginzburg-Landau (DGL) theory was developed as 
 the infrared effective theory of quarks and gluons \cite{Suzuki,Toki}. In 
the DGL theory, color confinement is  provided by  QCD-monopole condensation, 
which naturally appears by taking a particular gauge, 
the 't Hooft abelian gauge \cite{tHooft}.  
The recent lattice QCD simulations indicate that   
the abelian part of gluons plays dominant roles at low energy, 
while the non-diagonal gluons are irrelevant in the suitably chosen 
abelian gauge, which is called as the abelian dominance \cite{Suzuki,Ezawa}. 
Hence, the non-diagonal gluons can be neglected in the DGL theory.  
   The DGL theory was able to describe the linear potential 
 between a static quark-antiquark pair and at the same time the 
spontaneous chiral symmetry breaking \cite{Toki,Suganuma}. The crucial 
 role of monopole condensation on confinement and the abelian 
 dominance were clearly observed in the recent lattice 
 simulations \cite{Kronfeld}-\cite{STSM}. 
 The deconfinement phase transition at finite temperature is studied in terms 
of the effective 
 potential using the DGL theory \cite{Ichie,monden}. We are then able to 
extract 
 the surface tension from the bump structure of the effective potential 
 as a function of the QCD-monopole condensate \cite{Suganuma}.
 
  The bubble formation was discussed by Kajino in the ultra-relativistic 
heavy-ion collisions, where QGP of   order of the volume; $V \sim (10\,{\rm 
fm})^3 $ might be formed \cite{Kajino0}. On the way of 
hadronization,  the QGP phase may cool down to experience the mixed phase 
of the   QGP 
  and hadron, just as the case of big bang. In this case, 
  however, the size of each bubble is small and hence we ought to 
  take into account the finite volume effect for the 
  surface tension.
  
In ultra-relativistic heavy-ion collisions, we face at  another mixed 
 phase system. Just after the heavy-ion collisions, the color-flux tube 
 of finite size might be formed. The interior  of the flux tube is 
in the deconfinement phase while the outside is in the confinement phase. 
In stead of the temperature effect in the case of the 
ordinary mixed phase, the deconfinement region  is 
 produced by the color-electric field in the same manner as the 
vortex solution in superconductivity. We are also able to 
 investigate this case in the DGL theory. Thus, in the latter 
 coexistence system, the color electric field plays the essential 
 role. 
 
We  study in this paper the surface tensions of these mixed phase 
  systems in the DGL theory.  
We consider the flat surface case, where the two phases are 
  separated by one flat plane. We study also the various geometry 
  effect on the surface tension by considering various boundary shapes.

  This paper is organized as follows. In Sect. 2, we describe the  
  properties of the coexistence system using the DGL 
  theory at finite temperature. The surface tensions are extracted for the 
  flat surface case and also for the spherical case with various 
  radii. In Sect. 3, we study the vortex configurations in the DGL theory 
  at zero temperature.    We study the 
  surface tensions for the flat surface case and for the cylindrical 
  configuration case. Sect. 4 is devoted to the summary of the present study 
  and the discussion.}

\section{The surface tension at finite temperature}
  
{\indent

 The dual Ginzburg-Landau (DGL) theory is based  essentially on 
the abelian gauge fixing  {\it a la} 't Hooft  in non-abelian gauge theory 
as 
  QCD, which results in the appearance of the magnetic monopole.  In the 
  construction of the DGL theory,   only two assumptions are made 

i) The abelian dominance. The low energy phenomena or the long range behaviors of the quark-gluon 
  dynamics are dominated by the abelian part  and the 
  non-abelian part can be neglected. 
  
ii) The monopole condensation. The  magnetic monopole  is 
condensed due to the self-interaction, which lead to the 
color confinement \cite{Nielsen}. \\ 
These assumptions are strongly 
supported in the lattice gauge theory \cite{Kronfeld,Hioki}.

 }

 \subsection{DGL Effective Potential at Finite Temperature}
 
{\indent

  We consider  the deconfinement phase transition at finite 
  temperature. At the quenched level, the 
  DGL Lagrangian can be expressed in terms of the dual gauge fields 
  $\vec{B}_{\mu}$ and the  monopole fields $\chi_{a}$, 
\begin{equation}
{\cal L}_{DGL} = -\frac14 ({\partial_{\mu} \vec{B}_{\nu} -\partial_{\nu} 
\vec{B}_{\mu}})^2 + 
\sum^3_{a=1}\left[|(\partial_{\mu}-ig 
\vec{\epsilon}_{a}\cdot \vec{B}_{\mu})\chi_{a}|^2 -\lambda(  
|\chi_{a}|^2 - v^2)^2\right]
\end{equation}
 Here, $\vec{\epsilon}_{a}$, which are the root vectors of SU(3), 
 define the magnetic charges of the
 monopole fields. The dual gauge fields $\vec{B}_{\mu}$ 
 are  defined on the dual manifold of the Cartan subalgebra,  
 $\vec{B}_{\mu}=(B_{\mu}^3, B_{\mu}^8)$. The dual gauge coupling 
 constant $g$ satisfies the Dirac condition, $eg=4\pi$, with $e$ being the 
 gauge coupling constant \cite{Suganuma}.   
 At zero temperature, the monopole condensate is given as 
 $|\chi_{a}| = v$ at the mean field level for all $a$=1,2,3 due to 
 the self-interaction term. We know that the Lagrangian is the  dual 
 version of the Ginzburg-Landau theory of superconductor.  The main 
 difference from  the Ginzburg-Landau theory  is the existence of the 
 Weyl symmetry of SU(3) group \cite{STSM}, which stems from the SU(3)$_{c}$ 
 gauge of QCD. 
 
 We  derive the effective  
 potential at finite temperature by introducing the quadratic external 
 field, $J 
 \sum^{3}_{a=1}|\chi_{a}|^2$ \cite{Ichie}. The generating functional is 
 written as 
 \begin{eqnarray}
 Z[J]  &=&\langle 0|T e^{-i\int d^4 x  J 
\sum^{3}_{a=1}|\chi_{a}|^2}|0\rangle\nonumber\\  
       &=&\int {\it D\chi_{a}}{\it D \vec{B_{\mu}}}\exp \left(i \int  d^4 
       x ({\cal L}_{DGL}- J \sum^{3}_{a=1}|\chi_{a}|^2)\right) \equiv 
       e^{iW[J]}. 
 \end{eqnarray}
We rewrite the monopole field as 
 \begin{equation}
 \chi_{a}= (\bar{\chi}+ \tilde{\chi}_{a}) e^{i \xi _a},
 \end{equation}
 where the vacuum expectation value 
 $\bar{\chi}\equiv|\langle 0|\chi_{a}|0\rangle|$ is independent of 
 the index $a$ due to the Weyl symmetry. Here, the phase value 
 $\xi_{a}$ is absorbed in the dual gauge field $\vec{B}_{\mu}$ and 
 turns into its longitudinal component.  
 
 The DGL Lagrangian is written as 
 \begin{eqnarray}
 {\cal L}_{DGL}- J 
\sum^{3}_{a=1}|\chi_{a}|^2&=&-3\lambda(\bar{\chi}^2-v^2)^2-3J\bar{\chi}^2 
\nonumber\\
 & & -\frac14 (\partial_{\mu} \vec{B_{\nu}}-\partial_{\nu} 
 \vec{B_{\mu}})^2 + \frac12 
 m^2_{B}\vec{B^2_{\mu}}+\sum^3_{a=1}[(\partial_{\mu}\tilde{\chi}_{a})^2
 -m^2_{\chi}\tilde{\chi}_{a}^2] \nonumber\\
 & &+2\sum^3_{a=1}[-2\lambda \bar{\chi}(\bar{\chi}^2-v^2)
 -J\bar{\chi}]\tilde{\chi}_{a}\nonumber\\
 & &+\sum^3_{a=1}[g^2 
(\vec{\epsilon}_{a}\cdot\vec{B_{\mu}})^2(\tilde{\chi}_{a}^2
 +2\bar{\chi}\tilde{\chi}_{a})
 -\lambda(4\bar{\chi}\tilde{\chi}_{a}^3+\tilde{\chi}_{a}^4)],
 \end{eqnarray}
 where $m_{B} = \sqrt{3}g\bar{\chi}$ and $m_{\chi}=\sqrt{2\lambda(3 
 \bar{\chi}^2 -v^2)+J}=2\sqrt{\lambda}\bar{\chi}$.  Here, source $J$  
 is related to the mean field ${\bar{\chi}}$,  
 $ J= -2\lambda (\bar{\chi}^2-v^2)$,  through the condition that the 
 linear term of  $\tilde{\chi}_{a}$  vanishes in Eq.(4).

 We obtain the partition function $Z[J]$ by integrating over the quadratic part 
of 
 $\vec{B}_{\mu}$ and $ \tilde{\chi}_{a}$ up to the one-loop level. 
 The Legendre transform of the partition function gives 
 the effective action as 
 \begin{eqnarray}
 \Gamma(\bar{\chi}) &=& -i\,\ln\,Z[J]+\int  \,3J\bar{\chi}^2 d^4 x\nonumber\\
  &= &-\int 3\lambda(\bar{\chi}^2-v^2)^2  d^4 x + i\,\ln \, Det(i{\it 
D}^{-1}_{B}) 
  +3\frac 12 i\,\ln\, Det(i{\it D}^{-1}_{\chi}),\\
  & & {\it D}_{B} = 
  \left(g_{\mu\nu}-\frac{k_{\mu}k_{\nu}}{m^2_{B}}\right) 
\frac{i}{k^2-m^2_{B}+i\epsilon},\\
  & & {\it D}_{\chi} = \frac{-i}{k^2-m^2_{\chi}+i\epsilon}, 
 \end{eqnarray}
  where factor $3$ originates from the degrees of freedom of the QCD-monopole.
Using the imaginary time formalism, the effective potential 
$V_{{\rm eff}}$ at finite 
 temperature is obtained as \cite{Ichie},
 \begin{eqnarray}
 V_{{\rm eff}}(\bar{\chi};T) = 3\lambda(\bar{\chi}^2-v^2)^2 
 &+& 3\frac{T}{\pi^2}\int^{\infty}_{0}\,dk\,k^2\, 
\ln\,(1-e^{-\sqrt{k^2+m^2_{B}}/T})\nonumber\\
 &+& \frac32 \frac{T}{\pi^2}\int ^{\infty}_{0}\,dk\,k^2\, 
\ln\,(1-e^{-\sqrt{k^2+m^2_{\chi}}/T}), 
 \end{eqnarray}
 where the temperature dependent terms are derived from the functional 
 determinants in the momentum representation in Eq.(6) and Eq.(7). 
 
As for the parameters of the DGL theory for our investigation of the 
surface tension, we consider two cases. In one case, we take  
the parameter set so as to reproduce the inter-quark potential \cite{Toki}. 
We use this parameter set (case A) as the standard one to discuss all 
the features of deconfinement phase transition, since many 
non-perturbative phenomena have been discussed with this parameter 
set in the DGL theory \cite{Ichie,Toki,Suganuma}. 
This case corresponds 
to the dual  superconductor of type-II, in which $m_{\chi}>m_{B}$. 

In the other case (case B), we try to readjust the DGL parameters in 
order to take into account the recent lattice QCD results.
With lattice QCD simulations for the SU(3) gauge theory, 
the critical temperature has been studied in great detail. 
The continuum limit is given 
as \cite{Karsch0}, 
\[
T_{c}/\sqrt{k_{s}}= 0.625\pm 0.03
\]
with the expected systematic shift of 0.004 due to the infinite volume 
extrapolation of the critical couplings. With the use of the string 
tension, $\sqrt{k_{s}}=0.42$ GeV, we get a critical temperature of
$T_{c}\sim 0.26 $ GeV. 
Another constraint on the model parameters can be obtained from the 
scalar glueball mass, because the QCD-monopole in the DGL theory 
should appear as a scalar glueball with the mass 
about 1 - 2 GeV \cite{Suganuma}.
Very recently, Weingarten {\it et al.}\cite{Weingarten} worked out the 
enormous calculation on the mass spectrum and the decay pattern 
of low-lying glueballs using the lattice QCD, and concluded that 
$f_{0}(1710)$ is identical to the lowest scalar glueball with $J^{\pi}=0^{++}$. 
Therefore, we identify the QCD-monopole with this scalar-glueball 
candidate $f_{0}(1710)$, and adopt $m_{\chi}=1.71$ GeV for 
the QCD-monopole mass at $T=0$.
There is some indications on the type of the dual superconductor; the QCD vacuum 
corresponds to the case between type-I and 
type-II \cite{Maedan,Matsubara}. Since this part is not fully agreed 
upon by the specialists, we vary $m_{B}$ from 0.5 GeV to 2.0 GeV in 
order to study the parameter dependence of the surface tension. 
We take therefore,\\
case A: $\lambda = 25$, $v=0.126\,{\rm GeV}$ 
and $g=2.3$, which provides $m_{\chi}=1.26\, {\rm GeV}$ and 
$m_{B}=0.5 \,{\rm GeV}$. The string tension is fixed to $k_{s} = 1.0$ GeV/fm, 
which corresponds to $\sqrt{k_{s}}=0.444$ GeV.\\
Case B:  $m_{\chi}=1.71$ GeV, $\sqrt{k_{s}}=0.42$ GeV and $m_{B}$ is 
changed form 0.5 GeV to 2.0 GeV. \\

In order to get the feeling of the behavior of the free energy at 
finite temperature, we discuss here the results for case A. If we 
keep the coupling constants at finite temperature in the DGL 
Lagrangian as those fixed by the properties of the QCD vacuum at zero 
temperature, the critical temperature comes out to be too high, 
$T_{c}=0.49$ GeV, as compared to the lattice results. Since we work 
with the effective Lagrangian, we could vary any of the coupling 
constants with temperature. In fact, the asymptotic free property 
suggests the reduction of the coupling strength as the temperature 
increases. We follow the work of Ichie {\it et al.} \cite{Ichie} and 
take temperature dependence for  $\lambda$ as
\begin{equation}
\lambda (T) = \lambda \left( \frac{T_c-aT}{T_c}\right),
\end{equation}
and fix the parameter $a$ so as to reproduce $T_c=0.28\,{\rm GeV}$. 
They have found  $a=0.88$ and discussed the properties of various 
quantities at finite temperature. We show in Fig. 1 the behavior of the 
effective potential in this case. At small $T$, the absolute minimum 
in $V_{\rm eff}$ appears at finite $\bar{\chi}$, which is indicated by 
$\times$. The minimum point of $\bar{\chi}$ moves towards the smaller 
value with  
temperature. Around $T_{low}\sim 0.26
\, {\rm GeV}$, the local minimum 
appears at $\bar{\chi}=0$. This local minimum  becomes the absolute one 
at the critical temperature, 
$T_c\,=\,0.28\,{\rm GeV}$. At $T>T_{c}$, the absolute minimum stays  
at $\bar{\chi}=0$, which physically corresponds to the deconfinement 
phase.

We should make one more comment on the lower critical temperature 
$T_{low}$. It is defined as the temperature at which  
a local minimum of $V_{\rm eff}$ appears at the trivial vacuum 
$\bar{\chi}=0$. $T_{low}$ is exactly derived using  
the high temperature expansion \cite{Dolan} as

\begin{equation}
T_{low}=2v\left(\frac{6\lambda(T_{low})}{2\lambda(T_{low})+3g^2}\right)^{1/2}.
\end{equation}
We find $T_{low}=0.262\, {\rm GeV}$ for case A. We can check this formula through the agreement 
with the numerical calculations as shown in Fig.1.
We get similar results for case B for the free energy.

}

\subsection{The Surface Tension on the Flat Interface at Finite 
Temperature}

{\indent
 In general, a mixed phase  is realized for the case of
 the first-order phase transition. In the mixed phase, the  system 
 consists of two different phases, which are divided by an 
 interface. The free energy densities of both phases  
 are equal there.  The free energy of 
the mixed phase is different from the sum of the two free energies by the 
amount 
$F_{s}=\alpha S$, where $S$ is the area of the interface 
and $\alpha$ is the surface tension, which is the energy per unit 
interface.

 In the DGL theory, such a mixed phase consists of the QGP and the hadron 
 phases.  The free energy of the system is estimated using 
 the effective potential $V_{{\rm eff}}(\bar{\chi};T)$ as 
 \begin{equation}
  F(\bar{\chi};T) =\int d^3 x\left[\,3\,\left(\nabla
\bar{\chi}\right)^2 
  + V_{{\rm eff}}(\bar{\chi};T)\right],
 \end{equation}
 where $\bar{\chi}(x) $ is  the monopole condensate 
 in Eq.(3). The coefficient 3 
 of $(\nabla \bar{\chi})^2$ is the degree of  freedom of monopoles  
 originated from the Weyl symmetry for the SU(3) case.     

The surface tension on the planar interface can be estimated 
analytically, when the effective potential 
$V_{{\rm eff}}(\bar{\chi},T_{c})$ 
between two minima at $T_c$ is approximated by a sine curve \cite{Suganuma}. 
We show in Fig. 2 $V_{{\rm eff}}(\bar{\chi},T_{c})$ and the approximate 
sine curve,      
\begin{equation}
V_{{\rm sin}}(\bar{\chi})=\frac{ h}{2}
\left(1-\cos\frac{2\pi \bar{\chi}}{\bar{\chi}_H}\right),
\end{equation}
where $\bar{\chi}_H$ is the monopole condensate at the non-trivial 
local minimum of the effective potential, and 
$h$ is the barrier height of the effective potential 
between $\bar{\chi}=0$ and $\bar{\chi}=\bar{\chi}_H$.   
Here, we set $V_{{\rm eff}}(0;T_c)=V_{{\rm eff}}(\bar{\chi}_H;T_c)=0$. 
In this case, the system holds translational invariance in the 
direction parallel to the interface.
Hence, $\bar{\chi}$  depends only on $z$-coordinate perpendicular to its flat 
interface, and the surface tension is reduced to the one-dimensional 
integration over the energy density, 
\begin{equation}
\alpha_{\infty}[\bar{\chi}_{\rm sin}] 
=\int d z \left[\,3\,\left(\frac{d \bar{\chi}_{\rm 
sin}}{d z} \right)^2 
+ V_{{\rm sin}}(\bar{\chi}_{\rm sin})\right].
\end{equation}
Here, we use $\alpha_{\infty}$ because the flat surface case 
can be regarded as the hadron bubble with the infinite radius.
The field equation for $\bar{\chi}_{\rm sin}(z)$ is found to be the 
sine-Gordon (SG) equation, and can be solved analytically as \cite{Suganuma} 
\begin{equation}
\bar{\chi}_{{\rm sin}}(z)= \frac{2\sqrt 6}{3} 
\bar{\chi}_{H}\tan^{-1} e^{z/\varepsilon}
\end{equation}
with $\varepsilon = \frac{\sqrt 3}{\pi}\frac{\bar{\chi}_H}{\sqrt h}$. 
Here,  $2\varepsilon$ denotes the 
thickness of the boundary between the two phases. Substituting this 
solution 
into the expression for the surface tension in Eq.(13), 
\begin{equation}
\alpha_{\infty}=\frac{4\sqrt 3}{\pi}\sqrt{h} \bar{\chi}_H.
\end{equation}
Using the formula, we find $\alpha_{\infty}^{1/3}=114\,{\rm MeV}$ 
for case A. 

On the other hand,  the surface tension on the 
planar interface  is also calculated numerically  from $ V_{{\rm 
eff}}$ directly. The profile of $\bar{\chi}(z)$ is given by the 
differential equation as
\begin{equation}
\frac{d^2 \bar{\chi}}{dz^2}=\frac16\frac{\partial V_{{\rm eff}}}{\partial 
\bar{\chi}},
\end{equation} 
which is derived by the minimization condition of the free energy in 
Eq.(11). 
The surface tension is expressed as 
\begin{equation}
\alpha_{\infty} =\int d z \left[\,3\,\left(\frac{d \bar{\chi}}{d z} 
\right)^2 
+ V_{{\rm eff}}(\bar{\chi})\right].
\end{equation}

We solve Eq.(16) for $\bar{\chi}$ and find 
$\alpha_{\infty}^{1/3}=117\,{\rm MeV}$ for case A. This value is very 
close to the analytical result using Eq.(15). 
Hence, the use of the sine-Gordon (SG) kink solution in Eq.(14) is good for 
the surface tension.
This result seems to be consistent with the recent lattice 
results,  $\alpha_{\infty}^{1/3}= 80\,{\rm MeV}$ for $N_{t}=4$ 
and 73 MeV for $N_{t}=6$ \cite{iwasaki}.

We study also the parameter dependence of the surface tension. 
In the superconductivity, the Ginzburg-Landau (GL) parameter $\kappa$ 
is useful for categorizing the superconductor in terms of the magnetic 
response; the type I  and type II superconductors correspond to 
$\kappa < 1/\sqrt{2}$ and $\kappa > 1/\sqrt{2}$, respectively. This 
GL parameter is also meaningful for the dual superconductor picture 
of the QCD vacuum  given by $\kappa = m_{\chi}/\sqrt{2}m_{B}$ in 
the DGL theory. 
Some lattice simulations suggest that the QCD vacuum would correspond 
to the case near the border ($\kappa = 1/\sqrt{2}$) between  
type-I and type-II \cite{Matsubara}. 
In order to  understand  the behavior of the surface tension 
with respect to $\kappa$, we 
investigate various cases with  $0.61\,< \,\kappa \,< \,2.4$ in the 
case B. We list in Table 1 $m_{B}$,  $\kappa$, $g$, $\lambda$, 
$v$, $\lambda(T=T_{c})$ and $v(T=T_{c})$.  Here, we introduce also the 
temperature dependence of $v(T=T_{c})$, in the same manner as the 
temperature dependence of $\lambda(T=T_{c})$.  
$v(T=T_{c})$ is obtained by varying $v$ so as 
to reproduce  $T_{c}\sim 0.26$ GeV as $\lambda$ is fixed to the value 
at zero temperature. 

We show in Fig. 3  surface tensions $\alpha_{\infty}$, which are 
derived by using $\lambda(T=T_{c})$ and $v(T=T_{c})$ as a  function of 
the GL parameter $\kappa$. The surface tensions  decrease  
with $\kappa$. The surface tension is found to be $108 {\rm MeV} \le 
\alpha_{\infty}^{1/3} \le 241 {\rm MeV}$ for this range of $\kappa$. If 
the mass $m_{B}$ is small, the value of the surface tension is close 
to the value, $\alpha^{1/3}=80$ MeV, obtained in the lattice QCD simulations  
\cite{iwasaki}.  

Coleman discussed the scenario of the bubble formation in the early 
Universe \cite{Coleman}.  
In the thermal history of the Universe,  
as the temperature drops  gradually across the critical temperature 
 $T_{c}$ in the early Universe, the hadron bubble formation in the bulk QGP 
 phase begins at  $T_{c}$. It is considered that the phase transition from the 
QGP phase to the hadron phase is completed above $T_{low}$, 
because the latent heat turns into the shock wave emission from   the hadron 
bubble  in the beginning of the phase transition and its reheating effect  
suppresses the formation of  new  hadron bubbles. 

 
When $T>T_c$, the system is in the deconfinement phase. Near $T_c$, only 
extremely large hadron bubble can be formed, 
since the formation of a mixed phase necessarily needs the additional 
surface energy and the free energy of two bulk phases are almost 
equivalent \cite{Linde,Ichie4}. 
For $T_{low} < T < T_{c}$,  the finite radius $R(T)$ of the hadron  bubble 
is determined by the balance between the 
volume  and the surface energies of a bubble. 
There appears the energy difference between the interior and the exterior 
regions of 
the bubble, $\Delta V(T) =|\, V_{{\rm eff}}(0;T)-V_{{\rm eff}}(\bar{\chi}_H;T) \,|$,  
 as a source to  compensate the extra free energy density 
caused by the interface region. The sizable bubble configuration is 
allowed energetically \cite{Coleman,Linde}. 
From this consideration, we may assume that only 
the volume energy and the surface energy contribute to 
the free energy $F(T)$ as, 
\begin{equation} 
F(T)\simeq -\Delta V(T)\frac{4\pi}{3}R(T)^3 +\alpha_{\infty} 4\pi R(T)^2,
\end{equation}
where the source term is approximated by the surface tension of the infinite size 
bubble $\alpha_{\infty}$. In many studies, the surface tension of the finite 
size bubble $\alpha (R)$ is replaced by that of the infinite size 
bubble for simplicity,   
even for  the free energy of the finite size bubble. For a small 
bubble,  its free energy would be largely  different 
from $F(T)$  evaluated by Eq.(16).  
For the realistic estimation on the  free energy of the hadron bubble, 
we ought to consider the curvature effect on the  surface tension. 
}
 \subsection{The Surface Tension in the Hadron Bubble with 
 the Spherical Geometry}
{\indent

We  discuss, here,  the  radial dependence of the surface tension of 
hadron  droplets in the bulk QGP phase below $T_{c}$.
The free energy of a hadron  bubble in the  spherical geometry is 
given by 
\begin{equation}
  F(\bar{\chi};T) =\int dr 4\pi r^2\left\{\,3\,\left(\frac{d 
\bar{\chi}}{dr}\right)^2 
  + V_{{\rm eff}}(\bar{\chi};T)\right\},\\
\end{equation}
where   
$\bar{\chi}=\bar{\chi}(r)$  is the solution of the equation of motion 
in the spherical coordinate, 
\begin{equation}
\frac{d^2 \bar{\chi}}{dr^2}= -\frac{2}{r} \frac{d\bar{\chi}}{dr} 
+\frac16\frac{\partial V_{{\rm eff}}}{\partial \bar{\chi}}
\end{equation}
 with  the boundary conditions for the hadron bubble; $\bar{\chi}= 
\bar{\chi}_H $ 
at $r=0$ and $\bar{\chi}=0$ at $r \to \infty$. 

Generally, in 1+d dimensional scalar theory, the soliton equation does not 
necessarily have a nontrivial static solution because of the Derrick theorem
 \cite{Rajaraman}. 
The 1+1 dimensional field equation, which corresponds to Eq.(16), 
has always the solution to stabilize the free 
energy because the scale transformation as $r' = \eta r$ for 
$F_{\eta}(\bar{\chi},T)$ supplies a stable point on 
$\eta$,
\begin{eqnarray}
F_{\eta}(\bar{\chi};T)&=&\int_{-\infty}^{\infty}dr'\left[3\left(\frac{\partial 
}{\partial r'}\bar{\chi}(r')\right)^2\eta +V_{{\rm eff}}[\bar{\chi}(r')] 
\frac{1}{\eta}\right]\nonumber\\ 
&=&\eta \quad{\rm (Kinetic\,\, part)}+\frac{1}{\eta}\,\, {\rm 
(Potential\,\, part)}. 
\end{eqnarray} 
On the other hand, if d $\ge$ 2, there exist no nontrivial solutions, because the 
scale transformation of the radial coordinate does not make the free 
energy stable. For instance, in 1+3 dimensional case, which corresponds 
to Eq.(20). There is no stable point on $\eta$ 
in the free energy, $F_{\eta}(\bar{\chi};T)$ as 
\begin{eqnarray}
F_{\eta}(\bar{\chi};T)&=&\int_{0}^{\infty}4\pi dr'r'^2
\left[3\left(\frac{\partial }{\partial r'}\bar{\chi}(r')\right)^2 \frac{1}{\eta} 
+V_{{\rm eff}}[\bar{\chi}(r')]\frac{1}{\eta^3}\right]\nonumber\\  
&=&\frac{1}{\eta}\quad {\rm (Kinetic\,\, part)}+\frac{1}{\eta^3}\quad{\rm 
(Potential\,\, part)}.
\end{eqnarray}
Thus, there is no nontrivial static solution in the 1+3 dimensional scalar field 
theory, even if one resorts to any numerical methods.
However, our aim is not to find the static solution of Eq.(19), but  
is to study the properties and dynamics of the hadron bubbles, which appear 
due to the thermal fluctuation at the early Universe.
In general, the hadron bubbles are unstable against the scale 
transformation, so that they should expand or shrink as the time 
goes \cite{Ichie4}.

Here, we concentrate ourselves on the fixed size hadron bubble 
with the spherical geometry, and study its dynamical evolution in 
terms of the surface tension.  
When the bubble radius is large enough, the solution of Eq.(20) in 
spherical coordinate is reduced to that of the flat surface case, 
since we can neglect the first term in the right hand side at $R \to \infty$. 
We adopt the ``sine-Gordon (SG) kink form'' \cite{Ichie4} 
for the profile function of the monopole condensate $\bar{\chi}(r)$, 
\begin{equation}
\bar{\chi}(r)=\bar{\chi}_{H}\frac{\tan^{-1} e^{(R-r)/\rho}}
{\tan^{-1} e^{R/\rho}},
\end{equation}
which reproduces the exact form of Eq.(16) at $T_{c}$. 
As shown in Fig. 4, $R$ is the bubble radius  defined at the 
half of the monopole condensate $\bar{\chi}_H$ at which $V_{{\rm eff}}$ 
is absolute minimum and $2\rho$ 
corresponds to the thickness of the interface. We normalize the 
profile function so as to satisfy $\bar{\chi}(r=0)=\bar{\chi}_H$, 
which is the non-trivial minimum of the effective potential. 
In the supercooling environment, we take the minimization condition 
 by making  the free energy in 
Eq.(19) minimum for a fixed radius $R$ by varying $\rho$, in other words, by 
changing  the inclination of the SG-kink curve. We show in 
Fig. 5(a) and Fig. 5(b)  the profile of 
the monopole condensate $\bar{\chi}(r)$ and the free energy density at 
$T=0.2795\, {\rm GeV}$, as an example.  
As seen in Fig. 5(b), the free energy density is negative 
inside the bubble and positive at the bubble surface.  

Subsequently,  we plot the free energies for 
various radii $R$  in Fig. 5(c) at  $T=0.2795\, {\rm GeV}$ as the 
typical case. 
The free energy  has a peak at the critical bubble radius $R_c$, 
that  is, a branch radius for a bubble to shrink or grow 
during the phase transition. In general, there appears the critical 
radius $R_{c}$ at any temperature  from $T_{low}$ to $T_c$. 
One finds that Fig. 5(c) behaves in the same 
manner as the free energy in Eq.(18), although now the surface term is 
calculated more accurately. 

We estimate the surface tension at $R_{c}$, because the creation of 
the  hadron bubble with     $R_c$ is relevant in the hadronization 
process.  
Once $R_c$ is defined at various temperatures, the surface tension of 
the hadron bubble is 
derived by the substitution of $\alpha_{\infty}$ for $\alpha[R(T)]$ in 
Eq.(18),  
\begin{equation}
\alpha(R_c)=\frac{1}{4\pi R_c^2}\left\{F(\bar{\chi};T)\nonumber
+\Delta V(T) \frac{4\pi}{3}R_c^3\right\},
\label{eq:alpha}
\end{equation}
where the volume energy $\Delta V(T)$ does not dependent on $R_{c}$ 
at each temperature.


For case A at $0.278\,{\rm GeV} \le T \le 0.2799\,{\rm GeV}$, 
the radial dependence of the surface tension is shown at various 
temperatures in Fig. 6(a). One finds $\alpha (R_c)$ gradually approaches 
$\alpha_{\infty}$ with $R_c$.  On the other hand, $\alpha (R_c)$ 
increases rapidly with decreasing  $R_c$ below $R_c \sim 30  {\rm fm}$, 
although this behavior seems of inconsistent with Kajantie {\it et 
al.}'s  assertion \cite{Kajantie2} 
on the surface tension.

This behavior of the increase of $\alpha (R_c)$ at small $R_c$ 
can be explained from  the increase of the surface energy by the derivative term 
$\partial \bar{\chi}/\partial r$.
As shown in Fig. 6(b), the thickness of the bubble wall, which is about 
several  fm,  is found
 larger  than that in the lattice study \cite{Kajantie3}. 
 The behaviors 
of the free energy at various temperatures are shown in Fig. 6(c). 
They  are similar to the behaviors of the free energy in Eq.(18). 
However, quantitatively, there is a difference between the free 
energies with $\alpha_{\infty}$ and $\alpha (R_c)$  in the low energy 
region of interest for hadronization. 
As for case B,  
$\alpha [R(T)]$ behaves similarly as that of case A. 

We show the critical bubble 
radius $R_c$ during the phases transition at various temperatures in 
Fig. 7. 
Hadron bubbles would be born with various sizes corresponding to each 
temperature, and  
then  would grow up with the expansion of the Universe. The slight 
decrease of the temperature rapidly makes the possible bubble radius 
small. Since the bubble formation rate is expressed as $P\propto 
\exp \{-4\pi R_c(T)^3 h(T)/3T\}$, with $h$ being the barrier height between the 
false vacuum and the true one, the formation rate of  the  hadron 
bubble is drastically changed with temperature near $T_{c}$. 
} 

\section{The Surface tension on the flux tube }
 
{\indent

We  study here the surface tension in the color-flux tube system with 
considering the geometrical effect of the tube.   
In the  relativistic heavy-ion collisions, the huge size color 
 flux tube is  produced between  the target and the 
 projectile.
This scenario of forming huge size color flux tube has been pointed out by 
Ichie {\it et al.} \cite{ichie2}. 

In the previous section, the coexistence of the two phases is brought by 
the thermal effect.  On the other hand, the deconfinement phase is 
realized by the external color-electric field for the flux tube 
system.  Hence, to begin with, one should consider the free energy of 
the QCD system with the external color-field. 
In the case of the flux  tube,  the deconfinement phase contributes to the 
 equivalence of the free energies of each phase by the induced 
 color-magnetic current in the confinement phase. 
This configuration is 
easily inferred by  analogy with the coexisting system of the 
superconductor immersed in the magnetic field.  
The interior of the flux tube  corresponds  
 to the normal phase in the superconductor and its exterior 
 corresponds to the superconducting phase. 

First, we rewrite the interaction terms of 
 the monopoles ${\chi_{a}}$ and the dual gauge fields $\vec{B}_{\mu}$ 
 in the DGL Lagrangian into the simple form similar to the well-known 
 GL theory. The dual 
 gauge fields  $\vec{B}_{\mu}=(B^3_{\mu}, B^8_{\mu})$ are induced by  
the color current brought by each 
 quark-antiquark pair in the flux tube. 
 The quark charges $\vec{Q}_a=(Q_a^3,Q_a^8)$ determine the ratio of 
 $B_{\mu}^3$ on $B_{\mu}^8$ as 
\begin{eqnarray}
\vec{B}_{\mu} &=& \left( \frac12,\frac{1}{2\sqrt3}\right)B_{\mu} 
\qquad\qquad     {\rm for} \qquad
\vec{Q}_1=\left( \frac12,\frac{1}{2\sqrt3} \right)e,\nonumber\\
\vec{B}_{\mu} &=& \left(-\frac12,\frac{1}{2\sqrt3}\right)B_{\mu}
\qquad\quad     {\rm for} \qquad
\vec{Q}_2=\left(-\frac12,\frac{1}{2\sqrt3} \right)e,\\
\vec{B}_{\mu} &=& \left(0,      -\frac{1}{\sqrt3} \right)B_{\mu}
\qquad\qquad     {\rm for} \qquad
\vec{Q}_3=\left(      0,-\frac{1}{\sqrt3}  \right)e\nonumber.
\end{eqnarray}


As for $\vec{B}_{\mu}$, the relevant terms appearing in the DGL 
Lagrangian are $\vec{\epsilon}_a \cdot \vec{B}_{\mu}$, which have 
cyclic property on the index $a$,
 \begin{eqnarray}
 \vec{\epsilon}_1\cdot \vec{B}_{\mu}=0, \quad 
 \vec{\epsilon}_2\cdot \vec{B}_{\mu}&=&-\vec{\epsilon}_3\cdot 
 \vec{B}_{\mu}= -\frac12 {B}_{\mu}, 
 \qquad {\rm for}\quad Q_1-\bar{Q}_1,\nonumber \\
 \vec{\epsilon}_2\cdot \vec{B}_{\mu}=0, \quad 
 \vec{\epsilon}_3\cdot \vec{B}_{\mu}&=&-\vec{\epsilon}_1\cdot
 \vec{B}_{\mu}= -\frac12 {B}_{\mu},
 \qquad {\rm for}\quad Q_2-\bar{Q}_2, \\
 \vec{\epsilon}_3\cdot \vec{B}_{\mu}=0, \quad 
 \vec{\epsilon}_1\cdot \vec{B}_{\mu}&=&-\vec{\epsilon}_2\cdot 
 \vec{B}_{\mu}= -\frac12 {B}_{\mu},  
 \qquad {\rm for}\quad Q_3-\bar{Q}_3\nonumber.           
 \end{eqnarray} 
Here,  $\vec{\epsilon}$ is the root vector of SU(3),
\begin{equation}
\vec{\epsilon}_1=\left(-\frac12, \frac{\sqrt3}{2}\right),\,\,\,\,\,\,\,\,
\vec{\epsilon}_2=\left(-\frac12,-\frac{\sqrt3}{2}\right),\,\,\,\,\,\,\,\, 
\vec{\epsilon}_3=(1,0).
\end{equation}

It should be mentioned that two non-trivial field equations for $\chi_a$ 
 become equal, and there is one  trivial solution $|\chi|=v$, which is 
determined by the energy minimum condition. 
As a  result, we find the following relations of the 
components of  $\chi$ for each color charges of quarks,
 \begin{eqnarray}
|\chi_1|=v,\,\,\,\,\,\,\,\,|\chi_2|&=&|\chi_3|=|\chi|,
\qquad {\rm for}\quad Q_1-\bar{Q}_1,\nonumber\\
|\chi_2|=v,\,\,\,\,\,\,\,\,|\chi_3|&=&|\chi_1|=|\chi|,
\qquad {\rm for}\quad Q_2-\bar{Q}_2,\\
|\chi_3|=v,\,\,\,\,\,\,\,\,|\chi_1|&=&|\chi_2|=|\chi|,
\qquad {\rm for}\quad Q_3-\bar{Q}_3\nonumber.
\end{eqnarray}

 From this consideration, whichever
 the quark-antiquark pair we take, the DGL 
 Lagrangian (1) should be reduced to,
\begin{equation}
{\cal L}_{DGL} = -\frac13 \cdot \frac14 
(\partial_{\mu}B_{\nu}-\partial_{\nu}B_{\mu})^2 +2|(\partial_{\mu}-ig 
 B_{\mu})\chi|^2 - 2\lambda (|\chi|^2-v^2)^2.
\end{equation}
Furthermore, we rescale the fields and the coupling constants as 
\begin{equation}
B_{\mu} \to {\sqrt{3}}B_\mu,\qquad
\chi \to \frac{1}{\sqrt{2}}\chi,\qquad
\lambda \to  2 \lambda,\qquad
g \to  \frac{2}{\sqrt3}g,\qquad
v\to \frac{1}{\sqrt2} v,
\end{equation}
which reduce ${\cal L}_{DGL}$  into  
the GL type Lagrangian for the superconductor theory,    
\begin{equation}
{\cal L}_{DGL}= - \frac14 (\partial_{\mu}B_{\nu}-\partial_{\nu}B_{\mu})^2 
+|(\partial_{\mu}-i g B_{\mu })\chi|^2 
- \lambda (| \chi|^2-v^2)^2.
\end{equation}
From now on, we use this simple Lagrangian with the rescaled variables.
}
\subsection{The Surface Tension in the Mixed Phase with 
Color-Electric Field: \hspace{1cm} Flat Surface Case}

{\indent

To begin with, we consider the ordinary surface tension on the flat 
surface case, which can be regarded as the infinite radius limit for 
the cylindrical flux tube.

Let us 
set  the interface $yz$-plane ($x=0$).
Then, the system essentially depends only on the $x$-coordinate. 
The differential equations   are given by the minimum conditions of 
the free energy $F$;  $\delta F/\delta 
\chi^{\ast}=0$ and $\delta F/\delta {\bf B}= 0$. Taking ${\bf 
B}=(0,B(x),0)$ and 
${\bf \nabla}\times {\bf B}={\bf D}=(0,0,D(x))$, they are found to be 
\begin{eqnarray}
(i{\bf \nabla} + g {\bf B})^2 \chi -2\lambda v^2 \chi +2\lambda 
|\chi|^2 
\chi&=&0\\
{\nabla} \times {\bf D} + 2 g^2 |\chi|^2 {\bf B} &=& 0.
\end{eqnarray}
We impose the boundary condition as $\bar{\chi}=0$ and $D=E_{c}$ 
at $x \to -\infty$ (normal phase), and  $\bar{\chi}=v$ and $D=0$ 
at $x \to +\infty$ (superconducting phase).

In this mixed phase, the color flux is regarded as the external field 
for the 
nonperturbative vacuum so that we take into account the contribution 
from  the 
external field as $-\int E\, d D$. Here, the induced color-electric field 
$D$ is brought by the external 
field $E$. If $E$ is independent from $D$, the free energy of the system 
with the  
external fields $\tilde{f}$ is represented as $\tilde{f} =f_0 
+f-D E_c$,  
where $f$ is the free energy contributed by the change of the monopole 
condensate and $f_0$ is that of the whole system without the external 
fields. 
On the other hand, the free energy in the normal phase at $x \to 
-\infty$ is 
$\tilde{f_E}=f_0-E_c^2/2$ because $D=E_c$ and $-\int E\, d D
=-\int D\, d D|_{D=E_c}=-E_c^2/2$.  Here, $E_{c}$ denotes the 
critical color-electric field and is determined so as to 
fill the 
gap between the free energy densities for $\chi=0$ and 
$\chi=v$ as $E_{c}^{2}/2=\lambda v^{4}$. Hence, one can rewrite
$\tilde{f_E}=f_{0}- \lambda v^4$
The surface tension on the flat interface 
is given by the difference between $\tilde{f}$ 
and 
$\tilde{f_E}$,
\begin{eqnarray}
\sigma_{\infty}&=& 
\int ^{\infty}_{-\infty} dx (\tilde{f}-\tilde{f_E})\nonumber\\
&=& \int^{\infty}_{-\infty} dx 
\left[\frac{D^2}{2}+\left|\frac{d\chi}{dx}\right|^2-
g^2 B^2|\chi|^2-2\lambda v^2 |\chi|^2+\lambda |\chi|^4-E_c D 
+\frac{E_c^2}{2}\right]. 
\end{eqnarray}

In order to make these equations simple, it is convenient to introduce  
dimensionless variables expressed as 
$\hat{x}\equiv x/\delta$, $\hat{\chi}\equiv \chi/v$, 
$\hat{B}\equiv B/(E_c\delta)$ and 
$\hat{D}\equiv d\hat{B}/d\hat{x}= D/E_c$. The GL parameter  $\kappa$ is 
given  as
\begin{equation}
\kappa^2 \equiv \frac{\delta ^2}{\xi^2}
         = \frac12 \frac{\frac{1}{m_B^2}}{\frac{1}{m_{\chi}^2}}
         = \frac{\lambda}{g^2},
\end{equation}
 where $\delta$ is the penetration length of the color-electric field 
$\delta =1/m_B$ and  $\xi$ 
is the coherent length of the monopole condensate as 
$\xi=\sqrt{2}/m_{\chi}$. As the  definition,  $\kappa \leq 1/\sqrt 2$ 
corresponds to the type-I 
superconductor and  $\kappa \geq 1/\sqrt 2$ the type-II superconductor, 
respectively.

The field equations written with the dimensionless variables are  
\begin{eqnarray}
\hat{\chi}''&=&\kappa^2\left[\frac12 
\hat{B^2}\hat{\chi}-\hat{\chi}+\hat{\chi}^3\right]\\
\hat{B}''&=&\hat{\chi}^2\hat{B}.
\end{eqnarray}
The boundary conditions are given by $\hat{\chi}=0 $, $\hat{D}=1$ 
at $x \rightarrow  -\infty$ and $\hat{\chi}=1 $, $\hat{D}=0$ at 
$x \rightarrow  \infty$. The surface tension is simply expressed as 
\begin{equation}
\sigma_{\infty} = \int \delta 
E^2_c\left[\frac{2}{\kappa^2} \hat{\chi'}^2+
\hat{B'}(\hat{B'}-1)\right]d\hat x.
\end{equation}

We show in  Fig. 8 $\sigma_{\infty}^{1/3}$ versus $\kappa$ with the 
parameter set B. 
$\sigma_{\infty}^{1/3}$  moves 
from $ 180\,{\rm MeV}$ to $-310\,{\rm MeV}$ for $0.61 \le \kappa \le 
2.4$,  when  $m_{\chi}$ is fixed as $m_{\chi}=1.71$ GeV. 
At $\kappa=1/\sqrt 2$, the QCD vacuum is 
just at the border between type-I and type-II of the dual superconductor, 
and the surface tension is zero, which is consistent with the ordinal 
superconductivity.

It is a remarkable feature that the surface tension may be 
negative in the type-II dual superconductor. In this case, it is 
more stable that a thick flux tube splits into many flux tube units, 
because the larger interface 
of many flux tubes make the free energy of the whole system lower. 
Hence, the huge flux tube would be unstable against the splitting, 
which may provide the inhomogeneity in the QGP formation.
}

\subsection{The Surface Tension in the Flux Tube with Cylindrical Geometry}
 
{\indent
We consider here the color flux tube with cylindrical geometry, and 
study the finite size effect on the surface tension in the flux tube. 
Setting the flux tube along the $z$-axis, the color-electric fields 
${\bf D}(r) = 
(0,0,\frac{1}{r}\frac{d}{dr}(r\,B_{\phi}))$ is given by the vector 
potential $ {\bf B}=(0,B_{\phi}(r),0)$ using the cylindrical 
coordinates $(r, \phi, z)$. 
Taking  the monopole field as $\, \chi= \chi(r)e^{in\phi}$, 
the DGL equations (32) and (33) are expressed as 
\begin{eqnarray}
\frac{1}{\hat{r}}\frac{d}{d\hat{r}}\left(\hat{r}\frac{d\hat{\chi}}{d\hat{r}}\right)
&=&\kappa^2\left[\frac12\left(\frac{\sqrt2}{\kappa}\frac{n}{\hat{r}}-\hat{B}\right)^2\hat{\chi}
-\hat{\chi}+\hat{\chi}^3\right],\\
-\frac{d}{d\hat{r}}\frac{1}{\hat{r}}\frac{d}{d\hat{r}}\left(\hat{r}\hat{B}_{\phi}\right)&=&
 \left(\frac{\sqrt2}{\kappa}\frac{n}{\hat{r}}-\hat{B}\right)^2 
 \hat{\chi}^2
\end{eqnarray}
with the dimensionless variables as  
$\hat{r}=r/\delta$, $\hat{\chi}=\chi/v$, $\hat{B}=B/(E_c\delta)$ and 
$\hat{D}=\hat{B'}+\hat{B}/r =(B'+B/r)/E_c$.

Using the  cylindrical coordinate, the surface tension is written in the 
same manner as the flat interface case,
\begin{equation}
\sigma (R)=\frac{1}{2\pi R}\int \frac{E^2_c \delta 
}{2}\left[\frac{2}{\kappa^2}\hat{\chi}'^2
+\left(\frac{\sqrt2}{\kappa}\frac{n}{\hat{r}}-\hat{B}\right)^2 \hat{\chi}^2
-2\hat{\chi}^2+\hat{\chi}^4+(\hat{D}-1)^2\right]2\pi \hat{r}d\hat{r},
\end{equation}
where $R$ is the cylindrical radius of the flux tube.
Here, we define the flux tube boundary as the point satisfying, 
$\hat{\chi}=0.5$, that is, the monopole condensate $\chi$ takes the half 
value at the  boundary. 

The integer $n$ is the winding number related to the flux 
quantization,  
 and corresponds to the vortex number  in the superconductivity. 
We study the profile function $\chi(r)$ of monopole field with various 
number of $n$. 
In discussing the radius and the
multiple  winding number, it is meaningful only for type-I case, 
$\sigma (R) \,>\, 0$, where  one large size  flux tube is preferred. 
Otherwise, the negative surface tension leads the fragmentation of the 
huge flux tube  into many 
thinner flux units. The numerical results for $\chi(r)$ are shown in  
 Fig. 9(a) and (b), using the parameter set, $\kappa=0.61$, 
$m_{\chi}=1.71\, {\rm GeV}$, $m_B=2.0\,{\rm GeV}$ and  $v=0.163$ GeV, 
which corresponds to the type-I superconductor. In Fig. 9(b), 
we plot the numerical results on the 
flux tube radius $R$ and the winding number $n$. Here, we add the 
fitting curve, $R=\bar{r}\sqrt n$ with $\bar{r}= 0.21$ fm.  
Thus, $R$ increases proportionally with $\sqrt n$. 
From $n=1$ data in Fig. 9(a), 
 the radius of one flux tube is  estimated about $\bar{r}= 0.21$ fm. 
As the physical interpretation, the thick flux tube with the winding 
number $n$ can be regarded as the gathering system of $n$ flux tubes 
with the radius $\bar{r}$ and unite winding number, and therefore its 
cross section $S$ is given as $S= \pi R^2 \sim n \pi \bar{r}^2$. 
The surface tension $\sigma (R)$ goes down with the radius $R$ related to 
$n$, and   asymptotically  approaches to $\sigma_{\infty}$ in the flat 
surface case  at $R \rightarrow \infty$ 
as shown in Fig. 10. 

\section{Conclusion}
{\indent

 In the dual Ginzburg-Landau theory, we have studied the surface tensions 
in two different systems. One is the mixed phase near the critical 
temperature 
and the other is the coexistence system of two phases with 
external color-electric field at zero temperature. 

The surface tension at finite temperature is a relevant quantity 
for the argument of the bubble formation in the big bang scenario. In 
particular, the huge surface tension leads to strong inhomogeneity in the 
early Universe \cite{Witt}, which affects primordial  nucleosynthesis  
and the formation of the strange quark matter.  Not only for the 
cosmology, the surface tension is important also for the QGP creation 
in the ultra-relativistic heavy-ion collision. 
We have  investigated the surface tension for  the hadron  bubbles 
appearing in the QGP phase using the SG-kink curve 
ansatz for the  profile of the monopole condensate. 

In the flat surface case at the critical temperature,  
 we have first approximated the effective potential by 
a sine curve to investigate the mixed phase  analytically. 
We have thus obtained the useful analytical expressions for the surface 
tension and the thickness of the boundary surface in terms of the 
shape parameters of the effective potential. 
We have then calculated the surface tension 
numerically, and have verified the validity of the analytical formulae. 
The surface tension  $\alpha_\infty$  is obtained as 
$\alpha_\infty^{1/3}=117$MeV for case A, where case A corresponds
to a typical parameter set for type-II dual superconductor.

 We have studied  the surface tension with varying  
 the Ginzburg-Landau parameter  $\kappa=m_\chi/\sqrt{2}m_B$. The 
  value for the surface tension is 
$108 {\rm MeV} \le \alpha^{1/3}_{\infty} \le 241 {\rm MeV}$ for 
 $0.61 \le \kappa \le 2.4$. The surface tension 
tends to be small for large $\kappa$ case corresponding to 
the type-II dual  superconductor. These values should be compared 
with $\sigma^{1/3}=80$ MeV obtained by lattice QCD simulations.

We have studied also the finite size effect on the surface
tension in the hadron bubble with various radii for  $T_{low} < T < T_{c}$.
To this end, we have taken  an ansatz for the profile function $\chi(r)$ 
of the  monopole condensate in the radial coordinate. 
Using a fixed value of the bubble radius $R$, the thickness 
parameter $\rho$ is determined by the minimization condition of  the free 
energy. The free energy of the hadron bubble has a maximum  at the  
``critical'' radius $R_c$, which  is a branch radius of a bubble to shrink 
or  grow during the phase transition.
We have estimated then the surface tension at the critical radius $R_c$, 
$\alpha[R_c(T)]$, as the  relevant case for the created hadron 
 bubble \cite{Ichie4}. We have found that 
$\alpha (R_c)$ stay almost the value at $\alpha_\infty$
down to the radius $R_c \sim 30$fm, $\alpha (R_c)$ 
tends to increase  with decreasing $R_c$.

We have studied also the surface tension in  the mixed phase with the 
external color-electric
fields at zero temperature.
Such a mixed phase would be formed in the early stage of ultra-relativistic 
heavy-ion collisions \cite{ichie2}.
We have examined  the surface tension in the flat surface case, as the 
large flux tube limit, for the various Ginzburg-Landau parameter 
$\kappa =m_\chi/\sqrt{2}m_B $ near the boarder of  the type-I and type-II 
dual superconductor. 
We have found  
$-310 \,{\rm MeV} \le \sigma^{1/3}_\infty \le  180\,{\rm 
MeV}$ for $0.61 \le \kappa \le 
2.4$,  when  $m_{\chi}$ is fixed to $m_{\chi}=1.71$ GeV.  In the 
type-I dual superconductor with $\kappa\le 1/\sqrt{2}$, the surface 
tension  $\sigma_\infty$ is positive, so that many flux tubes tend to 
fuse into single huge flux tube \cite{ichie2}. On the other hand, in 
the type-II case with $\kappa \geq 1/\sqrt{2}$, $\sigma_\infty$ is 
negative, which means that the large size color flux 
tube is unstable against the separation  into many thinner  flux tubes. 
The surface tension 
vanishes, $\sigma_\infty=0$ just at the border  between
type-I and type-II dual superconductor ($\kappa=1/\sqrt 2$). 

Finally, 
we have formulated also the surface tension in a huge  color flux 
tube with the  large winding number $n$, which corresponds to the 
vortex number in the superconductivity. We find a simple relation 
between the radius $R$ and the winding number $n$, $R= \sqrt{n} 
\bar{r}$, where $\bar{r}$ is approximately the radius of the unite flux tube. 
The surface tension $\sigma (R)$ goes down to the asymptotic value 
$\sigma_\infty$
as $R \rightarrow \infty$. Quantitatively, $\sigma (R)$ can be 
regarded $\sigma_\infty$ for $R > 2{\rm fm}$.

\vspace{2cm}
H. M. acknowledges the hospitality of the Theory Group at RCNP-Osaka 
during her stay at RCNP, where this study has been worked out. 
We are grateful to  Prof. T.~Kajino for fruitful discussions on the QCD phase 
transition in the early Universe.

}


\newpage 
\begin{flushleft}
{\large\bf Figure Captions}
\end{flushleft}%
\newcommand{\namelistlabel}[1]{\mbox{#1}\hfil}
\newenvironment{namelist}[1]{%
\begin{list}{}
{\let\makelabel\namelistlabel
\settowidth{\labelwidth}{#1}
\setlength{\leftmargin}{1.1\labelwidth}}
}{%
\end{list}}
\begin{namelist}{Fig1xxxx}

\item[Fig.1] The effective potential $V_{\rm eff} (\chi,T)$
as the function  of the monopole condensate $\bar{\chi}$ at various temperatures 
for case A. 
The temperature dependence of the variable self-coupling $\lambda(T)$ 
is fixed so as to provide the critical temperature at $T_c=0.28 \,{\rm GeV}$. 
The absolute minimum at each temperature is denoted by the symbol $\times$. 
At $T_c=0.28\,{\rm GeV}$, there appear two absolute minima.

\item[Fig.2] The effective potential $V_{\rm eff}(\bar{\chi},T_c)$ 
at the critical temperature (denoted by solid curve) versus the 
monopole condensate $\bar{\chi}$ for case A. 
The dashed curve denotes an approximate sine curve.  
Here, $h$ corresponds to the barrier height on the transition between 
the hadron and the QGP phases, and $\bar{\chi}_H(T_c)$ denotes the value 
for the monopole condensate at the non-trivial minimum. 
The energy level is set so as to $V_{\rm eff}=0$ at $\bar{\chi}=0$. 

\item[Fig.3] The surface tension as a function of the  Ginzburg-Landau 
parameter $\kappa$ for case B.  The surface tension $\alpha_{\infty}^{1/3}$ 
decreases with $\kappa$. The solid curve is obtained by taking the 
linear temperature dependence for $\lambda(T)$ so as to reproduce the 
critical temperature at $T=0.26$ GeV, while  other parameters are 
fixed. As comparison, the dashed curve is the case with temperature 
dependent $v(T)$, while other parameters as $\lambda$ are fixed to the 
zero temperature value. 
 
\item[Fig.4] The profile of the hadron bubble in terms of 
the monopole condensate $\bar{\chi}(r)$ near the critical 
temperature $T_c=0.28 \,{\rm GeV}$  
for  case A. The solid curve denotes the exact solution calculated from 
$V_{\rm eff}(\bar{\chi},T)$. 
The dashed curve denotes the SG-kink ansatz,
2$\bar{\chi}_H \tan^{-1} e^{(R-r)/\rho}/\pi$, where
$\bar{\chi}_H$ is the monopole condensate at the non-trivial minimum 
of $V_{\rm eff}(\bar{\chi},T)$.

\item[Fig.5] (a) An example of the monopole profile function $\bar{\chi}(r)$,  
approximated by the SG-kink curve, as a function of the radial coordinate $r$. 
(b) The corresponding free energy density.
The width of the peak of the free energy density 
corresponds to the thickness of the bubble wall. 
(c) The total free energy of the hadron bubble as the function of 
various bubble radii $R$. The critical radius $R_c$ is defined at the 
saddle point of the free energy at each temperature. 
In this example for case A,
$R_c$ is around $26\,{\rm fm}$ at $T=0.2795 {\rm GeV}$.
 
\item[Fig.6] (a) The surface tension $\alpha^{1/3}$ as a function of the 
critical radius $R_c$ for various temperatures, 
0.278 GeV $\le T \le $ 0.28 GeV. The values added nearby the curve 
correspond to the temperatures in unit of GeV.
(b) The profiles of the monopole condensate at the critical radius $R_{c}$ 
at various temperature, $T=$ 0.278 GeV, 0.2798 GeV and 0.2799 GeV. 
The horizontal axis denotes the radial coordinate $r$. 
The dashed vertical lines show the thickness of the bubble wall 
at each temperature. 
(c) The solid curve denotes the free energy of the hadron bubble 
with the radius $R$ at various temperatures (0.2795 GeV, 0.2797 GeV 
and 0.2798 GeV) evaluated from the effective potential. 
The dashed curve denotes the free energy approximated by a sum of 
the volume and the surface terms, where $\alpha_{\infty}$ is used.
 
\item[Fig.7] The critical radius $R_c$ versus the temperature $T$ 
for case A. $R_c$ increases rapidly with $T$ and goes up to 
infinity at the critical temperature $T_c$. 

\item[Fig.8] The surface tension $\sigma_{\infty}$ for the flat 
interface as a function of the Ginzburg-Landau parameter $\kappa$. 
This range of $\kappa$ corresponds to 
$0.5\,{\rm GeV}\leq m_B \leq 2.0 \,{\rm GeV}$ and 
$m_{\chi}=1.71\, {\rm GeV}$ and $\sqrt{k_{s}}=0.42$ MeV. 
At $\kappa=1/\sqrt 2$, the QCD vacuum is just at the border 
between type-I and type-II in terms of the dual superconductor. 

\item[Fig.9] (a) The profile of the monopole condensate 
$\bar{\chi}(r)$  as a function of the radial coordinate $r$ 
for the flux-tube system with various winding numbers $n$. 
(b) Relation between the flux tube radius $R$ and $n$ (denoted by  
black dots) for $\kappa=0.61$ corresponding to 
the type-I dual superconductor. Here, the masses are 
$m_B=2.0\,{\rm GeV}$ and $m_{\chi}=1.71\, {\rm GeV} $. 
This behavior is well reproduced by the dotted curve 
$R=\bar{r} \sqrt n$, where $\bar{r}=0.21$ fm is 
almost the radius of the single flux tube with $n=1$. 

\item[Fig.10] The surface tension 
$\sigma^{1/3} (R)$  as a function of the radius $R$ of the flux 
tube. The used parameters are the same as in Fig.9.
Here, $\sigma_{\infty}$ is the value in the flat surface case. 

\end{namelist}

\newpage


\begin{thebibliography}{99}


\bibitem{Witt}E. Witten, Phys. Rev. {\bf D30} (1984) 272.

\bibitem{Kajino0}T. Kajino, Phys. Rev. Lett. {\bf 66} (1991) 125.  \\
T. Kajino, M. Orito, T. Yamamoto and H. Suganuma,  Proc. of Int. 
Workshop on ``Color Confinement and Hadrons'', edited by H. Toki 
{\it et al.}, (World Scientific, 1995)  263.

\bibitem{Sumi0}K. Sumiyoshi, T. Kajino, C.R. Alcock and G.J. Mathews, Phys. 
Rev. {\bf D42} (1990) 3963.

\bibitem{Karsch0}F. Karsch, Proc. of the Int. Workshop on 
``Color Confinement and Hadrons'', edited 
by H.Toki {\it et al.}, (World Scientific, 1995) 109. 

\bibitem{Karsch} J. Fingberg, U. Heller and F. Karsch, Nucl. Phys. {\bf 
B392} (1993) 493. \\ 
G.S. Bali, C. Schlichter and K. Schilling, 
Nucl. Phys. {\bf B} (Proc. Suppl.) {\bf 34} (1994) 216.

\bibitem{Kanaya}Y. Iwasaki, K. Kanaya, S. Kaya, S. Sakai and T. Yoshie,  
Nucl. Phys. {\bf B} (Proc. Suppl.) {\bf 47} (1996) 515.

\bibitem{Kajantie}K. Kajantie, L. K\"arkk\"ainen and K. Rummukainen, 
Phys. Lett. {\bf B223} (1989) 213; Nucl. Phys. {\bf B357} (1991) 693. 
 
\bibitem{Huang}S. Huang, J. Potvin, C. Rebbi and S. Sanielevici, Phys. Rev. 
{\bf D42} (1990) 2864. 

\bibitem{iwasaki} Y. Iwasaki, K. Kanaya, L. Karkkainen, K. Rummukainen
and T. Yoshie, Phys. Rev. {\bf D49}  (1994) 3540.\\
Y. Aoki and K. Kanaya, Phys. Rev. {\bf D50}  (1994) 6921. 

\bibitem{Ichie}H. Ichie, H. Suganuma and H. Toki, Phys.  Rev. {\bf D52}  
(1995) 2944.
        
\bibitem{Suzuki}T. Suzuki, Prog. Theor. Phys. {\bf 80}  (1988) 929.\\
S. Maedan and T. Suzuki, Prog. Theor. Phys. {\bf 81}  (1989) 229.

\bibitem{Toki} H. Suganuma, S. Sasaki and H. Toki, Nucl. Phys.{\bf B435} 
(1995) 207. \\
H. Toki, H. Suganuma and S. Sasaki, Nucl. Phys. {\bf A577} (1994) 353c. \\
H. Toki, S. Sasaki, H. Ichie, H. Monden and H. Suganuma, 
J. Korean Phys. Soc. (Proc. Suppl.) {\bf 29} (1996) S346.

\bibitem{tHooft}G. 't Hooft, Nucl. Phys. {\bf B190} (1981) 455.

\bibitem{Ezawa}Z.F. Ezawa and A. Iwazaki, Phys. Rev. {\bf D25} (1982) 2681; 
{\bf D26} (1982) 631. 
 
\bibitem{Sasaki}
S. Sasaki, H. Suganuma and H. Toki, Phys. Lett. {\bf B387} (1996) 145. \\
S. Sasaki, H. Suganuma and H. Toki, Prog. Theor. Phys. 
{\bf 94}  (1995) 373.  \\
S. Sasaki, H. Suganuma and H. Toki, Proc. of Int. Conf. on ``Baryons `95'', 
Santa Fe (World Scientific, 1996) 555.  

\bibitem{Suganuma} H. Suganuma, S. Sasaki, H. Toki and H. Ichie,  
Prog. Theor. Phys. (Suppl.) {\bf 120} (1995) 57.   \\
H. Suganuma, H. Ichie, H. Monden, S. Sasaki, M. Orito, T. Yamamoto and 
T. Kajino, Proc. of ``PANIC `96'', Williamsburg, 
May 1996 (World Scientific) in press.

\bibitem{Kronfeld}A.S. Kronfeld, M.L. Laursen, G. Schierholz and U.-J.Wiese, 
Phys. Lett. {\bf B198} (1987) 516.  \\
A.S. Kronfeld,  G. Schierholz and U.-J.Wiese, Nucl. Phys. {\bf B293} 
(1987) 461.   

\bibitem{Hioki}S. Hioki, S. Kitahara, S. Kiura, Y. Matsubara, O. Miyamura, 
S. Ohno and T. Suzuki, Phys. Lett. {\bf B272} (1991) 326. \\
S. Hioki, S. Kitahara, S. Ohno, T. Suzuki, Y. Matsubara and O. Miyamura, 
Phys. Lett. {\bf B285} (1992) 343.

\bibitem{Origuchi} O. Miyamura and S. Origuchi, Nucl. Phys. {\bf B} 
(Proc. Suppl.) {\bf 42} (1995) 538.  

\bibitem{STSM}H. Suganuma, A. Tanaka, S. Sasaki and O. Miyamura, 
Nucl. Phys. {\bf B} (Proc. Suppl.) {\bf 47} (1996) 302.

\bibitem{monden}H. Monden, T. Suzuki and Y. Matsubara, Phys. Lett. 
{\bf B294} (1992) 100.


\bibitem{Nielsen}H.B. Nielsen and P. Olesen, Nucl. Phys. {\bf B61} (1973) 
45.

\bibitem{Weingarten}J. Sexton, A. Vaccarino and D. Weingarten, 
Nucl. Phys. {\bf B} (Proc. Suppl.) {\bf 47} (1996) 128.

\bibitem{Maedan} S. Maedan, Y. Matsubara and T. Suzuki, Prog. Theor. Phys. 
{\bf 84} (1990) 130.

\bibitem{Matsubara} Y. Matsubara, S. Ejiri and T. Suzuki, Nucl. Phys. {\bf 
B} (Proc. Suppl.) {\bf 34} (1994) 176.

\bibitem{Dolan}L. Dolan and R. Jackiw,  Phys. Rev. {\bf D9}  (1974) 3320.

\bibitem{Coleman} S. Coleman, Phys. Rev. {\bf D15}  (1977) 2929; 
{\bf D16} (1977) 1248.

\bibitem{Linde} A.D. Linde, Nucl. Phys. {\bf B216}  (1983) 421.

\bibitem{Ichie4}H. Ichie, H. Monden, H. Suganuma and H. Toki, 
Proc. of Int. Symp. on ``Nuclear Reaction Dynamics of 
Nucleon-Hadron Manybody System'', Osaka, Dec. 1995 
(World Scientific, 1996) 246.  \\
H. Suganuma, H. Ichie, H. Toki and H. Monden, Proc. of Int. 
Workshop on ``Nuclear Reaction and Particle  Cosmo Physics'', Atami, 
Japan, Jan. 1996, in press. 


\bibitem{Rajaraman} R. Rajaraman, ``Solitons and Instantons'', 
North-Holland (1982) 1.

\bibitem{Kajantie2}K. Kajantie, J. Potvin and K. Rummukainen, Phys. Rev. 
{\bf D47}  (1993) 3079. 

\bibitem{Kajantie3}K. Kajantie, L. K\"arkk\"ainen and K. Rummukainen,
Nucl. Phys. {\bf B333}  (1990) 100.  \\
R. Brower, S. Huang, J. Potvin and C. Rebbi, Phys. Rev. {\bf D46} (1992) 2703. 

\bibitem{ichie2}
H. Ichie, H. Suganuma and  H. Toki, Phys. Rev. {\bf D54} (1996) 3382.


\end{thebibliography}
\end{document}